\documentclass{aa}  
\usepackage{textcomp}
\usepackage[utf8]{inputenc}
\usepackage{lscape}
\usepackage{lastpage}
\usepackage{amsmath,amssymb}
\usepackage{mathrsfs}	
\usepackage{multicol}

\usepackage{graphicx}
\usepackage{txfonts}
\usepackage{hyperref}
\hypersetup{
    colorlinks   = true, 
    urlcolor     = cyan, 
    linkcolor    = blue, 
    citecolor   = blue 
}
\usepackage{placeins}
\usepackage{xcolor}
\usepackage{caption}

\graphicspath{plots}

\newcommand{\rap}{\textacutedbl}
\newcommand{\lap}{\textgravedbl} 
{\newif\ifnotend
\notendtrue
\def\veclist{ABCDEFGHIJKLMNOPQRSTUVWXYZabcdefghijklmnopqrstuvwxyz.}
\def\top#1#2.{#1}
\def\tail#1#2.{#2.}
\loop\expandafter\xdef\csname bb\expandafter\top\veclist\endcsname%
{{\noexpand\bf\expandafter\top\veclist}}
\edef\veclist{\expandafter\tail\veclist}
\if\veclist.\notendfalse\fi\ifnotend\repeat}

\newcommand{\kpc}   {\,{\rm kpc}}
\newcommand{\pc}    {\,{\rm pc}}
\newcommand{\Mpc}   {\,{\rm Mpc}}
\newcommand{\Msun}  {\,M_{\odot}}
\newcommand{\Msunyr}  {\,{\rm M_{\odot} yr^{-1} }}

\newcommand{\Gyr}   {\,{\rm Gyr}}
\newcommand{\Myr}   {\,{\rm Myr}}
\newcommand{\yr}    {\,{\rm yr}}
\newcommand{\kms}   {\,{\rm km\,s^{-1}}}
\newcommand{\asec}  {\,{\rm arcsec}}
\newcommand{\asecq}  {\,{\rm arcsec^2}}
\newcommand{\amin}  {\,{\rm ^{\prime}}}

\newcommand{\magn}  {\,{\rm mag}}

\newcommand{\HI}    {{\rm H\textsc{i}}}

\newcommand{\Halpha}{{\rm H}_\alpha}

\newcommand{\rhost}     {\rho_{\star}}
\newcommand{\Sigmagas}  {\Sigma_{\rm gas}}

\newcommand{\rhogas}{\rho_{\rm gas}}

\newcommand{\hst}   {h_{\star}}
\newcommand{\zst}   {z_{\star}}
\newcommand{\hgas}  {h_{\rm gas}}

\newcommand{\Phitot}{\Phi_{\rm tot}}
\newcommand{\rs}    {r_s}
\newcommand{\rvir}  {r_{\rm vir}}
\newcommand{\vcmax} {v_{c,{\rm max}}}
\newcommand{\Rvcmax}{R_{v_{c,{\rm max}}}}
\newcommand{\vc}    {v_c}

\newcommand{\rhoc}  {\rho_{\rm c}}
\newcommand{\rhodm} {\rho_{\rm dm}}
\newcommand{\Mvir}  {M_{\rm vir}}
\newcommand{\Mdm}   {M_{\rm dm}}
\newcommand{\Mst}   {M_{\star}}
\newcommand{\Mgas}  {M_{\rm gas}}
\newcommand{\MHI}   {M_{\rm H\textsc{i}}}
\newcommand{\MHe}   {M_{\rm He}}

\newcommand{\Msat}  {M_{\rm sat}}
\newcommand{\Mhost} {M_{\rm host}}
\newcommand{\rt}    {r_{\rm t}}

\newcommand{\Ndm}   {N_{\rm dm}}
\newcommand{\Nst}   {N_{\star}}
\newcommand{\Ngas}  {N_{\rm gas}}

\newcommand{\ldm}   {l_{\rm dm}}
\newcommand{\lst}   {l_{\star}}
\newcommand{\lgas}  {l_{\rm gas}}

\newcommand{\tfric} {t_{\rm df}}
\newcommand{\tcross}{t_{\rm cross}}

\newcommand{\sech}  {{\rm sech}}

\newcommand{\MstS}    {M_{\star, \rm sat}}
\newcommand{\Reff}    {R_{\rm e, sat}}
\newcommand{\dlosS}   {\Sigma_{\star, {\rm sat}}}

\newcommand{\MdmS}    {M_{\rm dm, sat}}

\newcommand{\admS}    {a_{\rm sat}}
\newcommand{\rhostS}  {\rho_{\star, {\rm sat}}}
\newcommand{\PhistS}  {\Phi_{\star, {\rm sat}}}

\newcommand{\Lz}    {L_z}
\newcommand{\mpart} {m_{\rm part}}

\newcommand{\sigmaR}    {\sigma_R}
\newcommand{\sigmaphi}  {\sigma_{\phi}}
\newcommand{\sigmaz}    {\sigma_z}
\newcommand{\phimean}   {v_\phi}

\newcommand{\xyz}   {(x,y,z)}
\newcommand{\vxyz}  {(v_x,v_y,v_z)}
\newcommand{\vx}    {v_x}

\newcommand{\dd}{\text{d}}
\newcommand{\DD}{\partial}

\newcommand{\Sersic}{S\'{e}rsic}
\newcommand{\Ho}    {H_0}

\begin{document} 

\title{
Beyond the surface: hydrodynamical $N$-body simulations of the interacting dwarf galaxies NGC 5238 and UGC 8760
}
\titlerunning{N-body models of NGC 5238 and UGC 8760}

\author{
R. Pascale \inst{1} \thanks{\email{raffaele.pascale@inaf.it}} \and
F. Annibali \inst{1} \and 
M. Tosi \inst{1} \and
C. Nipoti \inst{2} \and
F. Marinacci \inst{2} \and
M. Bellazzini \inst{1} \and 
J. M. Cannon \inst{3} \and
L. Schisgal \inst{3} \and
E. Sacchi \inst{4} \and  
F. Calura \inst{1} 
}
\institute{
INAF - Osservatorio di Astrofisica e Scienza dello Spazio di Bologna, Via Gobetti 93/3, 40129 Bologna, Italy 
\and
Dipartimento di Fisica e Astronomia \lap Augusto Righi\rap, Università di Bologna, via Piero Gobetti 93/2, I-40129 Bologna, Italy 
\and
Department of Physics and Astronomy, Macalester College, 1600 Grand Avenue, Saint Paul, MN 55105, USA \and
Leibniz-Institut für Astrophysik Potsdam, An der Sternwarte 16, 14482 Potsdam, Germany }
\authorrunning{Pascale et al.}
\date{Received ...; accepted ...}
 
\abstract
{
From deep imaging data obtained with the Large Binocular Telescope as part of the Smallest Scale of Hierarchy Survey (SSH), we have discovered low-surface brightness tidal features around NGC 5238 and UGC 8760, two nearby and relatively isolated dwarf galaxies with stellar masses of approximately $10^8\Msun$ and $2\times10^7\Msun$, respectively. In this study, we present detailed hydrodynamical $N$-body simulations that explain the observed faint substructures as the outcome of interactions between the dwarf galaxies and smaller satellite systems.  We show that the asymmetric stellar distribution of NGC 5238 and the low-luminosity substructures observed to the northeast of UGC 8760 can be well attributed to recent interactions with smaller galaxies, each with a stellar mass roughly a few $10^5\Msun$, 50 times less massive than their respective hosts. In the simulations, these satellites have stellar and dark-matter masses consistent with the ones predicted by $\Lambda$CDM cosmology and share properties similar to those of local dwarf galaxies with similar stellar masses. The satellite-to-main galaxy mass ratio is approximately 1:10 in both cases. This satellite population aligns closely with predictions from cosmological simulations in terms of the number and mass relative to the host galaxy mass.
}


\maketitle

\section{Introduction}
\label{sec:int}

\begin{figure*}
    \centering
    \includegraphics[width=1\hsize]{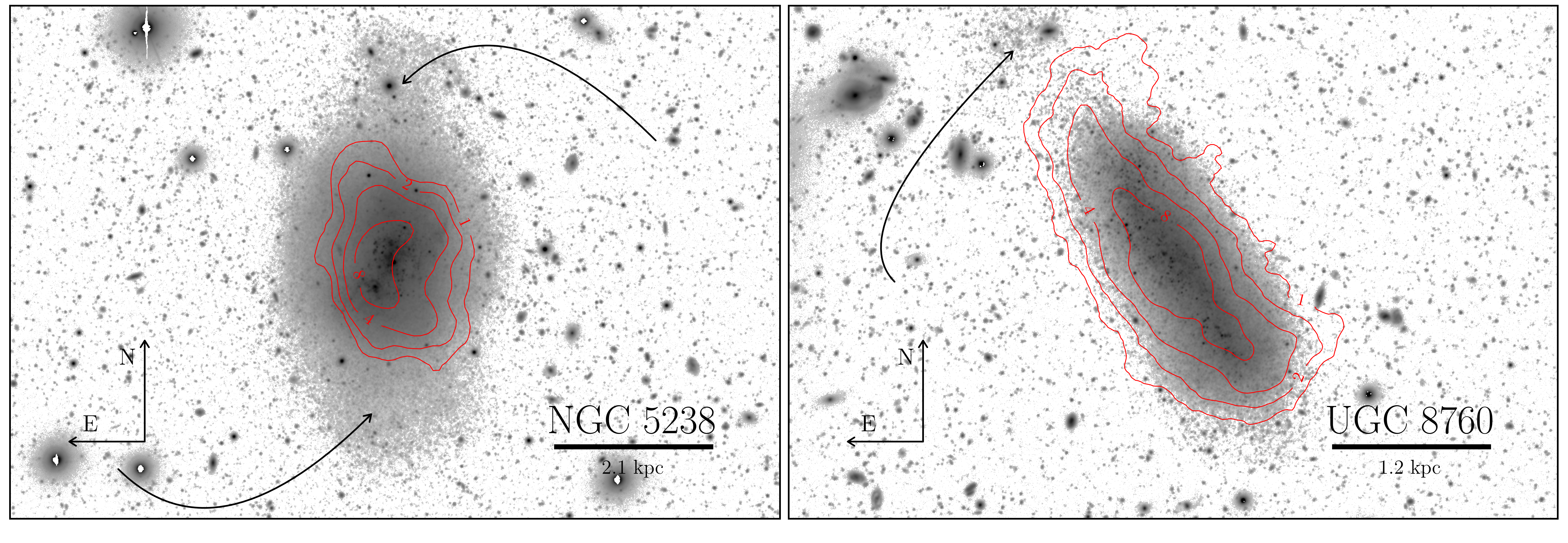}
    \caption{LBT images of NGC 5238 and UGC 8760. Left-hand panel: LBT g-band image of NGC 5238. The arrows show the positions of the northern and southern overdensities of stars while the red contours show lines of constant $\HI$ emission from VLA observations (\citealt{Cannon2016}) at the adopted distance. Right-hand panel: same as the left-hand panel but for UGC 8760. Here, the arrow shows the position of the northern overdensity of stars, while the VLA observations are from \citet{Begum2008}. In both panels, the red contours correspond to $1, 2, 4, 8 \Msun\pc^{-2}$. North is up, east is to the left.}
    \label{fig:galaxies}
\end{figure*}

The current $\Lambda$CDM cosmological model predicts that galaxies grow in size and mass thanks to mergers with smaller systems \cite[e.g.][]{White1978,White1991,Springel2005b,Genel2014,Planck2020}. According to cosmological numerical simulations, this pattern extends down to the smallest mass scales \citep{Diemand2008,Wheeler2015}, where dwarf galaxies form by accreting even smaller systems. However, direct observational evidence for satellites around dwarf hosts is still poor \citep[see][and references therein]{Annibali2022}, due to the challenges associated with detecting very low-surface brightness companions. Additionally, dwarf satellites located in the least massive dark matter sub-halos are expected to be totally dark \citep{Sawala2015,Read2017}. 


Among the population of dwarfs, it is expected from simulations that the ones living in low-density environments may have been unaffected by potential disruptive interactions with larger systems and that they may be still surrounded by a relatively large population of smaller galaxies and stellar streams \citep{Donghia2008,Deason2015,Wheeler2015,Jahn2019}. Indeed, once a group of dwarfs has been accreted into a more massive host, like the Milky Way, it will be  disassembled by tidal forces wiping out any evidence for coherent structure in about $\sim$5-6 Gyr \citep{Deason2015}. Therefore, isolated dwarfs are the most promising targets for studying the hierarchical assembling process at the lowest mass scales. 

The {\bf S}mallest {\bf S}cale of {\bf H}ierarchy \citep[SSH;][]{Annibali2020} survey is a recently completed long-term program at the Large Binocular Telescope (LBT) aimed at identifying merger signatures, typically visible as low surface brightness distortions or tidal stellar features, around a sample of 45 late-type dwarf galaxies in the local ($D\leq11\Mpc$) Universe. The deep optical SSH images are complemented with ancillary $\HI$ data from the literature providing information on the gas morphology and kinematics. To mention some of these surveys: the Little Things survey \citep{Hunter2012}, with the Very Large Array (VLA); the WHISP survey \citep{Swaters2002}, with the Westerbork Synthesis Radio Telescope; the VLA-ANGST survey \citep{Ott2012}; the FIGGS survey \citep{Begum2008}, with the Giant Metrewave Radio Telescope (GMRT). The ultimate goal of SSH is to characterize the frequency and properties of accretion events around local dwarf systems with different masses and in different environments; this is accomplished by means of state-of-the art hydrodynamical $N$-body simulations \citep[e.g.][]{Pascale2021,Pascale2022} capable of reconstructing the dwarfs' merger histories from the observed properties of gas and stars. 

An exceptional example of a multiple-merging dwarf galaxy, observed as a pilot study case for SSH, is the extremely metal poor \citep[$\sim$3\% solar oxygen abundance,][]{Pustilnik2005,Izotov2009}, gas-rich dwarf DDO~68 \citep{Annibali2016}, with a stellar mass of $\sim10^8\Msun$ \citep{Sacchi2016}. DDO~68 resides in the nearby Lynx-Cancer Void \citep{Pustilnik2011}, and it exhibits a bright, $10\kpc$-long stellar stream that crosses the galaxy from south to north, and a second, less luminous stellar stream to the northwest, detected through deep LBT data \citep{Annibali2019b}. By means of hydrodynamical $N$-body simulations, \cite{Pascale2022} showed that the system's structural and kinematic properties are well reproduced by an interaction between DDO~68 and two smaller satellite galaxies about 150 times and 20 times less massive than DDO~68. However, a third system discovered in VLA $\HI$ data \citep{Cannon2014}, and recently resolved into individual stars by \cite{Annibali2023}, has been proposed to be interacting with DDO~68 as well, making it a unique case of a small dwarf caught in the process of accreting three smaller satellites. Systematic searches for merger signatures around large samples of dwarf galaxies, like those tackled by SSH and other similar surveys \citep{Carlin2016,Higgs2016}, may provide insights into the environmental dependence of accretion events around the smallest dark matter halos in the present-day Universe.

In this paper, we present two new compelling cases of dwarf-dwarf mergers identified through visual inspection of the SSH LBT images, namely NGC~5238 and UGC~8760, for which we managed to successfully reproduce the stellar and gaseous observed properties through hydrodynamical $N$-body simulations.  Simulations of galaxies and mergers, despite being idealized setups, provide a complementary perspective to cosmological simulations, offering detailed insights into the processes shaping individual galaxies.  For instance, employing higher spatial and time resolutions than in cosmological simulations enables a more detailed analysis of the internal dynamics and interactions within individual systems, facilitating direct comparisons with observational data for specific objects. Moreover, unlike the intricate dynamics of cosmological simulations (wherein concurrent processes complicate the identification of key factors molding a galaxy), isolating and studying individual mergers offer a clearer perspective on galaxy formation and evolution. As an example, studying the merger of galaxies with distinct properties can shed light on the origin of structures like streams or tidal tails. 
Here, we aim to explain, within a cosmologically motivated framework, the observed substructures in UGC 8760 and NGC 5238 as the outcome of interactions with satellite galaxies. The simulations showcase potential scenarios of interactions, offering plausible explanations based on observational evidence. However it is important to bear in mind that there is no claim of uniqueness in the proposed solutions. Indeed, while simulations of interacting galaxies allow for a more detailed exploration of specific problems, the parameter space to be explored is so vast that, along with computational limitations and potential simplifications in the simulation setup, it becomes nearly impossible to definitively exclude other possible scenarios.

\begin{figure*}
    \centering
    \includegraphics[width=1\hsize]{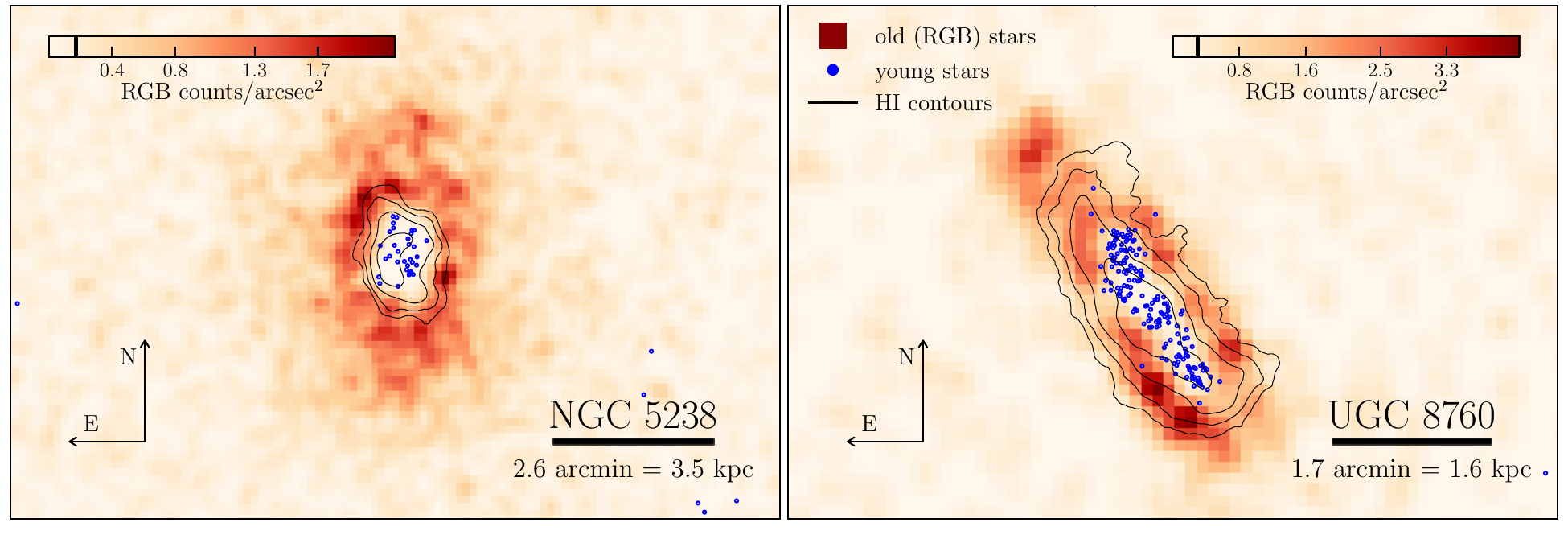}
    \caption{Spatial distribution of old (age$>2\Gyr$) RGB stars and young (age$\lesssim$100 Myr) bright blue stars in NGC 5238 (left) and UGC~8760 (right). RGB maps are background-subtracted, and only pixels above a level of 3 times the background standard deviation (given by the vertical black segment in the colorbars) have been set to non-zero to mask density peaks due to red background galaxies. The blue points are individual young stars. $\HI$ contours at $1, 2, 4, 8\Msun\pc^{-2}$ (same as in Fig.~\ref{fig:galaxies}) are superimposed to the star count maps. North is up, east is to the left.}
\label{fig:rgbstars}
\end{figure*}

NGC~5238 (also known as UGC 8565, MRK~1479, and IZw~64) is a gas-rich \citep[$\MHI\simeq(2.8-2.9)\times10^7\Msun$,][]{Cannon2016,Calzetti2015} dwarf irregular galaxy at the periphery of the Canes Venatici I group, at a distance of $D=4.51\pm0.04\Mpc$ \citep{Tully2009} from the Earth. According to its low tidal index $\Theta=-0.4$ (a quantifiable metric of the density contrast from massive neighboring galaxies; \citealt{Karachentsev2013}), NGC 5238 is fairly isolated, more than $1\Mpc$ from any known massive galaxy. NGC 5238 was observed in the context of several surveys, such as the H$\alpha$ 11HUGS survey \citep{Kennicutt2008}, the GALEX ultraviolet (UV) imaging survey \citep{Lee2011}, the infrared {\it Spitzer} Local Volume Legacy (LVL) survey \citep{Dale2009}, the HST optical ANGST survey \citep{Dalcanton2009} and the HST UV LEGUS survey \citep{Calzetti2015}. With an oxygen abundance less than 20\% solar \citep{Lira2007} and a stellar mass of $\Mst\sim10^8\Msun$ \citep{Cook2014,Cannon2016}, the galaxy is compatible with the mass-metallicity relation derived for star forming dwarfs \citep{Berg2012}. The star formation rate density ($0.53\times10^{-3} \Msun \yr^{-1}\kpc^{-2}$), derived from modeling the color-magnitude diagram (CMD) of resolved stars in the galaxy, is consistent with the value inferred from the $\Halpha$ flux \citep[$0.65\times10^{-3} \Msun \yr^{-1}\kpc^{-2}$; ][]{Cignoni2019}. The VLA $\HI$ images acquired by \cite{Cannon2016} reveal an irregular and asymmetric morphology in the outermost detected regions, and significant velocity asymmetries are present throughout the disc. Nevertheless, \cite{Cannon2016} do not report evidence for an ongoing interaction considering the sensitivity and angular resolution of the VLA observations. The region of current star formation, as traced by the UV and H$\alpha$ emission, exhibits a $\sim300\pc$ offset with respect to $\HI$ peak, a behaviour already observed in other dwarf galaxies \citep[e.g.][]{Roychowdhury2009}. 

The dwarf irregular UGC 8760 (or DDO 183), at a distance of $D=3.2\pm0.1\Mpc$ \citep{Dalcanton2009,Tully2009,Tully2013}, is another fairly isolated galaxy \citep{Karachentsev2013} and resides in the outer zone of the Ophiuchus–Sagittarius–Capricornus void \citep{Pustilnik2019}. As for NGC 5238, multi-wavelength data of UGC 8760 are available from various surveys (e.g., LVL, 11HUGS, ANGST, VLA-ANGST, and FIGGS). Stellar mass estimates derived from different methods are in the range $\Mst=(2-4)\times10^7\Msun$ \citep{Johnson2013,Cook2014}, while the $\HI$ mass from FIGGS and VLA-ANGST data is $\MHI \lesssim2.6\times10^7\Msun$ \citep{Begum2008,Ott2012}. The $\HI$ disk extends for a diameter of $\sim4\kpc$ \citep{Roychowdhury2017} and it shows coherent rotation with some level of distortion in the inner regions \citep{Begum2008,Ott2012}. According to the resolved-star CMD modeling and the integrated UV flux, the galaxy has been actively forming stars with an average rate of $\sim4\times10^{-3}\Msunyr$ over the last hundred $\Myr$ \citep{Lee2009,Johnson2013}. To our knowledge, no spectroscopic metallicity estimates are available for this galaxy.

This paper is structured as follows: In Section~\ref{sec:ssh_data}, we present the new LBT data where signatures of merger events have been identified. Section~\ref{sec:models} provides a brief description of the simulation setup, while Section~\ref{sec:res} presents and discusses our results. Finally, in Section~\ref{sec:conc}, we draw our conclusions.

\section{LBT data}
\label{sec:ssh_data}

NGC~5238 and UGC~8760 were observed with the Large Binocular Camera (LBC) on the LBT in the $g$ and $r$ bands as part of the SSH survey. Fig.~\ref{fig:galaxies} shows $8\amin\times5.3\amin$ (left panel) and $6.4\amin\times4.2\amin$ (right panel) portions of the larger 23$\amin\times$23$\amin$ imaged fields of view, with the $\HI$ contours from VLA data from \citet[NGC~5238]{Cannon2014} and from \citet[UGC~8760]{Begum2008} superimposed. In the following, we briefly summarize the main properties of the stellar populations detected in the two galaxies and their associated substructures. A detailed and thorough analysis is presented in \cite{Sacchi2024}. The deep LBT data reveal the presence of extended, low surface brightness stellar structures such as a northern ``umbrella'' and a southern plume in the case of NGC~5238, and an extended, disrupted-satellite feature to the northeast of UGC~8760. 

These stellar features are also visible in the resolved-star maps of Fig.~\ref{fig:rgbstars}, where we plot the distribution of both old red giant branch (RGB) stars and young blue bright stars. A detailed description of the point-source photometry performed on the LBT images and of the selection of RGB and young blue stars based on their location in the $r$ versus $g$-$r$ CMD are provided in \cite{Annibali2020} and \cite{Annibali2022b}. Here, it is sufficient to recall that RGB counts trace old stars with ages between $\sim$2 and $13\Gyr$, while the bluest and brightest counts in the CMD correspond to stars typically younger than about a hundred $\Myr$. These maps provide therefore a powerful tool to disentangle short-lasting, irregular structures due to active star-formation from distortions in the old stellar component, most likely of tidal origin.

Indeed, in the case of NGC~5238, the ``umbrella''-like feature detected in the deep LBT images coincides with a similar structure in the RGB count map, indicating that it cannot be accounted for by recent and offset bursts of star formation, but that it can be either the remnant of an accreted companion or a distortion of NGC~5238 of external origin.  More in general, we note that the old stellar halo of NGC~5238 extends for a diameter of $\sim360\asec$, or $\sim8\kpc$, well beyond the outer displayed $1 \Msun\pc^{-2}$ $\HI$ contour. Similar to NGC~5238, the elongated northeast feature detected in UGC~8760 coincides with a clear RGB star overdensity. However, no extended stellar halo is detected in this galaxy. In both NGC~5238 and UGC~8760, stars younger than $\sim100\Myr$ are confined to the innermost galaxy regions, producing significant stellar crowding and preventing the detection of faint RGB stars \citep{Cignoni2010,Okamoto2015,Cignoni2019,Higgs2021}. 




\section{Setting up the simulations}
\label{sec:models}

In this Section, we explain how we determined the relevant parameters of the galaxy models used to generate the initial conditions (ICs) of the simulations for NGC 5238, UGC 8760, and their associated satellites. We will explore the specific properties of the host galaxies and their satellite counterparts in Sections~\ref{subsec:hosts} and \ref{subsec:sat}, respectively. Additionally, Section~\ref{sec:ICs} will provide a detailed description of our simulations setup.

\subsection{Hosts} 
\label{subsec:hosts}

\subsubsection{The galaxy models}
\label{subsubsec:models}

Consistently with their classification (SABdm), NGC 5238 and UGC 8760 are modeled as dwarf disc galaxies with a dominant dark-matter halo, and thick stellar and gas discs. The dark matter halo follows the \citet[][$\gamma=0$]{Dehnen1993} profile

\begin{equation}\label{for:dm}
    \rhodm(r) = \frac{\Mdm}{2\pi}\frac{a}{(r + a)^4},
\end{equation}
with $\Mdm$ and $a$ the total dark matter mass and the halo scale length, respectively. Here $r$ is the three dimensional distance from the halo center. Although the NFW profile is cosmologically motivated, we prefer to use a Dehnen profile because: {\it i)} it has finite mass, being more practical to sample the halo particle phase-space positions, and {\it ii)} we expect dwarf galaxies to be filled with cores of constant dark matter densities rather than cusps (\citealt{Bullock2017} and reference therein). 

The density distribution of the stellar disc is
\begin{equation}\label{for:dstar}
    \rhost(R,z) = \frac{\Mst}{4\pi\zst\hst^2}e^{-\frac{R}{\hst}}\sech^2\biggl(\frac{z}{\zst}\biggr).
\end{equation}
In the above equation, $R$ and $z$ are the cylindrical coordinates of a reference frame whose $z$-axis is the disc's symmetry axis, $\Mst$ is the total stellar mass, $\hst$ is the stellar disc scale length and $\zst$ is the stellar disc scale height. The gas surface density follows the exponential profile
\begin{equation}\label{for:dgas}
    \Sigmagas(R) = \frac{\Mgas}{2\pi\hgas^2}e^{-\frac{R}{\hgas}},
\end{equation}
where $\hgas$ is the gas disc scale length and $\Mgas$ is the total gas mass, given summing up the mass contributions of neutral hydrogen ($\MHI$) and helium ($\MHe$):
\begin{equation}\label{for:mgas}
    \Mgas = \MHI + \MHe.
\end{equation}
We impose the vertical hydrostatic equilibrium
\begin{equation}\label{for:hydro}
 \frac{\DD P}{\DD z}\frac{1}{\rhogas} = -\frac{\DD\Phitot}{\DD z},
\end{equation}
to compute the gas disc scale height, and
\begin{equation}\label{for:rhogas}
 \Sigmagas(R)=\int_{-\infty}^{+\infty}\rhogas(R,z)\dd z.
\end{equation}
In equations (\ref{for:hydro}) and (\ref{for:rhogas}), $\Phitot$ is the total gravitational potential, $\rhogas$ is the gas density, $P=\rhogas(\gamma-1)e$ its pressure, $\gamma=5/3$ the adiabatic index and $e$ its specific internal energy, which we assume to be constant throughout the galaxy.

We assume that the distribution function (DF) of all the collisionless particles depends on the specific energy $E$ and the specific third component of the angular momentum $\Lz$ only, which makes all the mixed moments of the DF and the mean velocities in the vertical and radial directions null \citep{BinneyTremaine2008}. The non-null elements of the velocity ellipsoid (the vertical and radial velocity dispersions $\sigma\equiv\sigmaz=\sigmaR$ and the velocity dispersion in the streaming direction $\sigmaphi$) are determined from the axisymmetric Jeans equations. We sample the halo particle velocities in Maxwellian approximation: in the vertical and radial directions, the velocity distributions are Gaussians with null mean and standard deviation equal to $\sigma$, while, in the streaming direction, the velocity distribution is a Gaussian with non-null mean $\phimean$ and $\sqrt{\sigmaphi^2-\phimean^2}$ standard deviation (i.e. the halo is allowed to rotate). The stellar disc particles are sampled in Maxwellian approximation, but the streaming velocity is set by the epicyclic approximation. The only non-null gas velocity component is the azimuthal one, found imposing the stationary Euler equation (equation 21 of \citealt{Springel2005}). The total potential is found iteratively, while the potentials of the discs are computed numerically from a discretized mass distribution following equations (\ref{for:dstar}) and (\ref{for:dgas}) with a hierarchical multipole expansion based on a tree code (details in \citealt{Springel2005}).

\begin{figure}
    \centering
    \includegraphics[width=1\hsize]{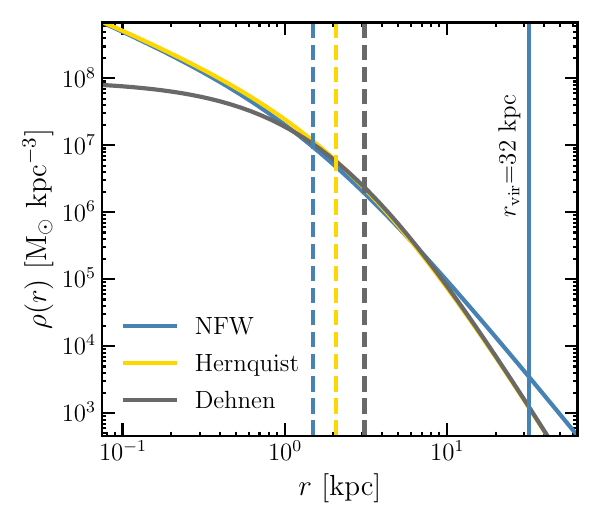}
    \caption{Dark matter density distribution of NGC 5238 used in the simulations described via three different density models (different colors are associated to the three dark-matter models as indicated in the legend). The different models are linked via equations (\ref{for:onetoone}, see text for details). The vertical dashed lines show the NFW, Hernquist and Dehnen scale radii using the same color coding, while the vertical blue solid line is the NFW virial radius.}
    \label{fig:halos}
\end{figure}

\subsubsection{Fixing the free parameters}
\label{subsubsec:param}

In the context of simulating interactions between galaxies, such as those observed in UGC 8760 and NGC 5238, careful consideration is given to setting: (i) the mentioned galaxy models, (ii) the orbital parameters defining the merger. For instance, we do not have a formal understanding of whether the properties of the galaxies post-interaction will necessarily resemble those pre-interaction. However, in the simulations, we set the latter to reproduce the former. Nonetheless, the selection is guided by logical assumptions and observational evidence. For instance, as we discuss in Section~\ref{sec:ICs} and confirm in Section~\ref{sec:res}, the dynamics of the interactions represented by the simulations we propose is not destructive for the host galaxies. Also, the simulations lack star formation (an additional mechanism for gas consumption), stellar feedback, and, as we will see in Section~\ref{subsec:sat}, the baryonic contributions of the satellites to the total baryonic mass are minimal. With these considerations in mind, we adopt the strategy to select the parameters of the pre-interaction galaxies based on observations, even though observations refer to a post-interaction scenario.

To determine the parameters of the Dehnen halo (\ref{for:dm}) we link them to those of a cosmologically motivated NFW \citet*[hereafter NFW]{NavarroFrenkWhite1996} model. In this way, when fixing the NFW model we fix the Dehnen profile. A NFW model is given by the density distribution
\begin{equation}
    \rho(r) = \frac{\delta\rhoc}{\frac{r}{\rs}\bigl(\frac{r}{\rs}+1\bigr)^2},
\end{equation}
with $\rhoc$ the present day critical density of the Universe (we assume $\Ho=67.4\kms\Mpc^{-1}$; \citealt{Planck2020}) and 
\begin{equation}\label{for:delta}
    \delta = \frac{200c^3}{3[\ln(1+c)-c/(c+1)]}.
\end{equation} 
The halo virial radius $\rvir$ is defined as the distance from the center where the average halo density equals 200 times $\rhoc$, and it is linked to the concentration $c$ by $c\equiv\rvir/\rs$. The halo virial mass $\Mvir$ is, then, the mass enclosed within $\rvir$. We determine the parameters of the Dehnen halo (\ref{for:dm}) using equations
\begin{equation}\label{for:onetoone}
\left\{
    \begin{array}{lr}
        \Mdm  = \Mvir, \\ 
        a=\rs\sqrt{\frac{8}{9}[\ln(c + 1)-c/(c + 1)]}. \\
\end{array}
\right.
\end{equation}
The above relations establish, analitically, the link between the parameters of a given Dehnen profile and the parameters of a general NFW model. A similar approach has been used by \cite{Springel2005} to link the parameters of the \cite{Hernquist1990} model to that of a cosmologically motivated NFW profile. As in \cite{Springel2005}, we impose that the Dehnen total mass equals the NFW virial mass (first equation of \ref{for:onetoone}), while we require that the Dehnen profile shares the same outer normalization of the Hernquist profile  where both decrease as $r^{-4}$. As an example, Fig.~\ref{fig:halos} shows the halo density distribution of NGC 5238 obtained with the three different halo models (NFW, Hernquist and Dehnen) linked via equations (\ref{for:onetoone}). Here, the overall agreement among the different models can be visually observed. In the simulations we then adopt the Dehnen halo.

In a NFW model, the maximum circular velocity is reached at a distance
\begin{equation}
    \Rvcmax = k\rs,
\end{equation}
where it assumes the value 
\begin{equation}
    \vcmax^2 \equiv \vc^2(\Rvcmax) = 4\pi G\rs^2 \delta\rhoc\frac{k}{(1+k)^2},
\end{equation}
with $\delta$ as in equation (\ref{for:delta}) and $k=2.162$ a dimensionless constant\footnote{ The value of $k$ is found imposing $\frac{\dd\vc}{\dd r}=0$.}. The quantities $\Mvir$ and $\rs$ (thus, $\Mdm$ and $a$) have been determined imposing that $\vcmax$ and $\Rvcmax$ approximately match the deprojected maximum circular velocity and the galactocentric distance from the $\HI$ kinematics.

\begin{table*}
    \centering
    \caption{Parameters used to generate the ICs of the galaxy models used in the simulations.} 
    \label{tab:params}
    \renewcommand{\arraystretch}{1.25}
    \begin{tabular}{lcccccccc}
        \hline
        \hline
        Galaxy  & $\Mdm$ [$10^9\Msun$] & $a$ [kpc] & $\Mst$ [$10^7\Msun$] & ($\hst$, $\zst$) [kpc] & ($\Reff$ [kpc], $m$) & $\MHI$ [$10^7\Msun$] & $\hgas$ [kpc] \\       \hline
        \hline
        NGC 5238    &  3.44 &   2.07    &   8.9     &   (0.52, 0.052)    &   -     &   2.74    &   0.52  \\
        Satellite   &  0.3  &   0.99    &   0.02    &   (0.15, 0.015)    &  -   &   0.015   &   0.3   \\
        UGC 8760    &  1.91 &   1.87    &   2      &   (0.275, 0.028)     &    -  &   2.64   &   0.55  \\
        Satellite   &  0.15  &   0.5    &   0.06 &   -   &  (0.15, 0.75)  &   -   &   - \\
        \hline
        \hline
    \end{tabular}
    \tablefoot{$\Mdm$ and $a$ are the dark-matter mass and scale radius (equations~\ref{for:dm} and~\ref{for:sat8760dm}); if the stellar component is a disc, then $\Mst$, $\hst$ and $\zst$ are the total stellar mass, scale length and scale height of the disc, respectively (equation~\ref{for:dstar}); otherwise $\Mst$, $\Reff$ and $m$ are the total stellar mass, the effective radius and the \Sersic\,index of the corresponding \Sersic\,model, respectively (equation~\ref{for:sat8760st}). $\hgas$ is the gas disc scale length (equation~\ref{for:dgas}), $\MHI$ is the $\HI$ mass and the Helium mass ($\MHe$) is always $0.35\MHI$.}
\end{table*}

\begin{table*}
    \centering
    \caption{Main simulation parameters.}\label{tab:params2} 
    \renewcommand{\arraystretch}{1.25}
    \begin{tabular}{lcccccc}
        \hline
        \hline
        Galaxy  & $\Ndm$ & $\ldm$ [pc] & $\Nst$ & $\lst$ [pc] & $\Ngas$ & $\lgas$ [pc] \\ 
        \hline
        \hline
        NGC 5238    &  4825618 &   10    &   127142     &   5    &   52857    &   5  \\
        Satellite   &  419752  &   10    &   571    &   10    &   571   &   5   \\
        UGC 8760    &  5328651 &   20    &   57142      &   15   &   101714   &   10  \\
        Satellite   &  429210   &   10 &  5143 &   10    &   -  & - \\
        \hline
        \hline
    \end{tabular}
    \tablefoot{$\Ndm$, $\Nst$ and $\Ngas$: number of dark-matter, stellar and gas particles in the simulations. Note that $\Ngas$ is number of gas particles in the ICs of the equilibrium simulations, and that the use of mesh refinement changes the number of gas resolution element throughout the simulations. $\ldm$, $\lst$ and $\lgas$: dark matter, stellar and gas softening lengths, respectively.}
\end{table*}

Below we list the specific parameters adopted to set up the galaxies' models.

NGC 5238. We impose $\vcmax=31.5\kms$ at $\Rvcmax=3.15\kpc$, larger than $\Rvcmax=1.3\kpc$ measured by \cite{Cannon2016}. This choice is motivated by the fact that the value of $\Rvcmax$ from \cite{Cannon2016} implies an unreasonably high concentration ($c\simeq40$). Therefore, we prefer the value of $\Rvcmax=3.15\kpc$, which ensures a more cosmologically motivated concentration ($c\simeq20$) while keeping a dynamical mass (Dehnen dark matter halo, stars and gas, see below) of $2.94\times10^8\Msun$ within a $1.3\kpc$ radius, in very good agreement with a mass of  $3\times10^8\Msun$ from \cite{Cannon2016}. To compute the deprojected maximum circular velocity, \cite{Cannon2016} derived an inclination of $i=50^{\circ}\pm5^{\circ}$, in agreement with the previous estimate by \cite{Lee2011} of $i=55^{\circ}$. 
For the stellar mass, we adopt $\Mst=0.89\times10^7\Msun$, taken from \cite{Cook2014} but scaled for the closer distance ($\sim4.5\Mpc$ versus $5.2\Mpc$) assumed in this work. This value is smaller than the $1.4\times10^8\Msun$ stellar mass of \cite{Calzetti2015}. We assume $\MHI=2.74\times10^7\Msun$, within the errorbars as in \citet[$(2.8\pm0.3)\times10^7\Msun$]{Cannon2016}, plus a correction for helium of $\MHe=0.35\MHI$ \citep[e.g.;][]{Leroy2008,Bolatto2013,Hardwick2022}. The total gas mass is $\Mgas=3.7\times10^7\Msun$, as computed by \cite{Cannon2016}. To determine $\hst$ and $\hgas$, we fitted both the RGB star count profile derived from the LEGUS catalogs  \footnote{The LEGUS photometric catalogs are available through the public site https://archive.stsci.edu/prepds/legus/dataproducts-public.html} and the $\HI$ surface density profile from \cite{Cannon2016} with exponential disc models. 
Although the fit provides $\hgas\simeq\hst=0.47\kpc$, for our simulations we adopt a larger value of $\hst=\hgas=0.52\kpc$ to keep the stellar disc stable against bar formation when evolved in isolation (see Section \ref{sec:ICs}). For the same reason, we prefer the smaller stellar mass from \cite{Cook2014} to that provided by \cite{Calzetti2015}. From a qualitative point of view, decreasing the stellar mass and/or increasing the scale radius, decreases the stellar disc surface density. This favors the stability of the stellar disc against gravity. Indeed, a razor-thin disc is locally unstable against gravity when its \cite{Toomre1964} parameter $Q\equiv\frac{\sigmaR\nu}{3.36\pi G\Sigma_\star}<1$, with $\nu$ the local epicyclic frequency and $\Sigma_\star$ the stellar surface density.
    
UGC 8760. We require $\vcmax=25\kms$ at $3\kpc$, consistent with the velocity peak observed in the $\HI$ velocity field map of UGC 8760 from the VLA-ANGST survey, assuming an inclination $i\simeq67^{\circ}\pm3^{\circ}$ \citep{Begum2008}. The implied $\Mdm$ and $a$ of the corresponding Dehnen halo are listed in Table~\ref{tab:params}, alongside all the other relevant parameters used to generate the galaxy model.
For the stellar disc, we adopt a mass of $\Mst=2\times10^7\Msun$ from \cite{Cook2014}. The stellar disc scale length is $0.275\kpc$, taken from \cite{Makarov2017} who modeled the galaxy B,V and I surface brightness emission via the same exponential model as in equation (\ref{for:dstar}). We set $\MHI=2.64\times10^7\Msun$, larger than the value from \citet[$1.86\times10^7 M_{\odot}$]{Georgiev2010}, but consistent within the errors with the estimate from \citet[$2.59\times10^7 M_{\odot}$]{Begum2008}. We assume that the helium mass is, again, $\MHe=0.35\MHI=0.93\times10^7\Msun$. 

\begin{figure*}
    \centering
    \includegraphics[width=1\hsize]{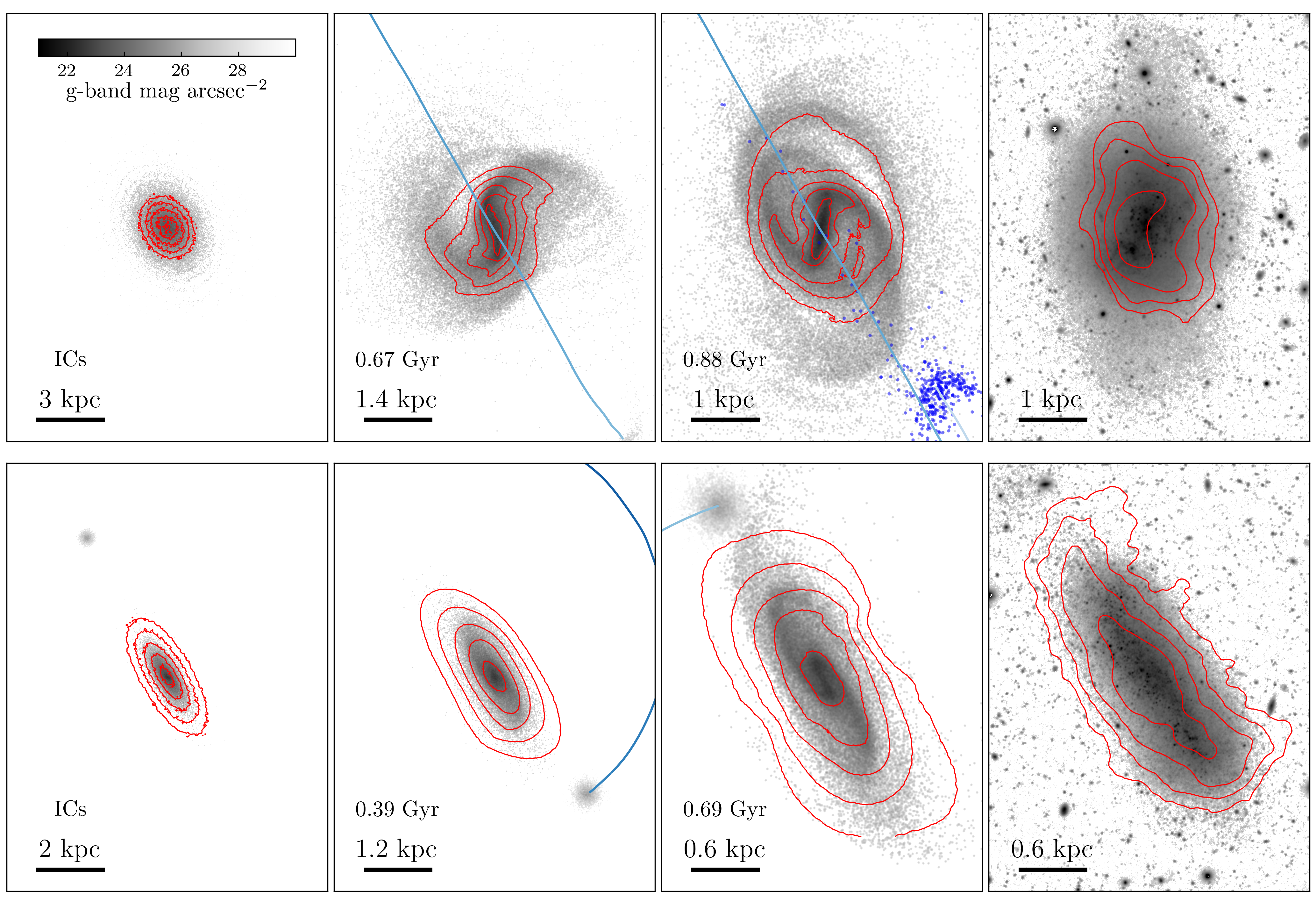}
    \caption{Comparison between simulations and observations. First row of panels: stellar surface brightness maps from the reference simulation of NGC 5238 and the corresponding LBT observation. The left panel shows the ICs of the simulation, while the middle panels are subsequent snapshots (as reported in each panel). The third panel corresponds to the time when the simulation reproduces the data, shown for comparison in the rightmost panel. In the simulated maps, the pixel size is $0.225\asec\times0.225\asec$ (as for the LBC camera), and the maps have been convolved with a Gaussian beam of $1\asec$ FWHM to match the typical seeing during observations. To generate the maps, we adopted constant mass-to-light ratios and we set the surface brightness limit to $30\magn\asec^{-2}$, to match the depth of the data. Blue lines show the orbit of the satellite, obtained by joining the centers of mass of its dark matter halo computed with the shrinking sphere method \citep{Power2003}. The color gradient indicates the direction of the orbit itself: the satellite moves from where the orbit is darker to where it is lighter. The red contours are levels of iso-$\HI$ column density, specifically $2^k\Msun\pc^{-2}$, with $k=0,..,3$. Second row of panels: same as the top row but for the UGC 8760 system. 
    For details on the simulation setup see Section~\ref{sec:ICs} and Tables \ref{tab:params} and \ref{tab:params2}.
    In the top row, third panel, we highlight in blue the stellar particles belonging to the satellite of NGC 5238 as a comparison (see text for details).}\label{fig:8760mock}
\end{figure*}

\subsection{Satellites}
\label{subsec:sat}

As discussed in Section~\ref{sec:ssh_data}, the stellar populations observed within the substructures of NGC 5238 and UGC 8760 suggest that the hypothesis of their formation through an asymmetric event of star formation is implausible. On the contrary, an external origin appears more probable. The stellar masses of NGC 5238 and UGC 8760 range between $2\times10^7\Msun$ and nearly $2\times10^8\Msun$. According to \cite{Dooley2017}, for galaxies with these masses there is a significant probability, 50\% for the former and up to 80\% for the latter, of hosting at least one satellite with a stellar mass of $10^5\Msun$ at most. This provides the basis for our approach to model the properties of the satellite galaxies in the simulations.

\subsubsection{The satellite of NGC 5238}
\label{subsubsec:5238}

The satellite galaxy of NGC 5238 is a scaled-down version of its host: a disc galaxy with a dominant Dehnen dark matter halo, an exponential thick stellar disc, and an exponential gas disc. As previously mentioned, we expect dwarf galaxies with stellar mass similar to NGC 5238 or UGC 8760 to have, at most, one satellite with a few $10^5\Msun$ \citep{Dooley2017}. Thus, we fix the satellite stellar mass to $\Mst=2\times10^5\Msun$, while $\hst=0.15\kpc$. The expected virial mass (from the stellar-to-halo mass relation in \citealt{Read2017}) and concentration (from the virial mass-to-concentration relation in  \citealt{MunozCuartas2011}) are, respectively, $\Mvir=3\times10^8\Msun$ and $c=19.5$, which we use, in combination with equations (\ref{for:onetoone}) to fix $\Mdm$ and $a$. The gas mass is $\Mgas=10\Mst$ \citep{Geha2006,CFN2020}, while $\hgas=2\hst$.


\subsubsection{The satellite of UGC 8760}
\label{subsubsec:8760}

In UGC 8760, the roundish shape of the northern overdensity of stars, along with the absence of $\HI$ emission at that location, motivates us to model its satellite as a gas-free, dark-matter dominated, spheroid rather than a small disk galaxy. We assume that dark-matter density of such a satellite follows the Hernquist model
\begin{equation}\label{for:sat8760dm}
    \rhodm(r) = \frac{\Mdm a}{2\pi r}\frac{1}{(r+a)^3}.
\end{equation}
In equation~(\ref{for:sat8760dm}), $\Mdm$ and $a$ are, respectively, the total mass of the dark matter halo and its scale radius. The stellar component is a spherically symmetric \cite{Sersic1968} model with surface density
\begin{equation}\label{for:sat8760st}
    \dlosS(R) = \frac{b^{2m}}{2\pi m\Gamma(2m)}\frac{\MstS}{\Reff^2}\exp\biggl[-b\biggl(\frac{R}{\Reff}\biggr)^{\frac{1}{m}}\biggr],
\end{equation}
with $\Gamma$ the gamma function, $\MstS$ the total stellar mass, $\Reff$ the effective radius (the projected distance from the center that contains half of the total mass), $m$ the \Sersic\, index, and $b=b(m)$ as in equation 18 of \cite{Ciotti1999b},  which ensures that $\int_0^{\Reff}2\pi\dlosS(R)R\dd R = \MstS/2$.

We assume that the halo and the stellar velocity distributions are isotropic, and we sample their particle positions and velocities directly from the corresponding ergodic DFs \citep{BinneyTremaine2008}
\begin{equation}
    f_{i,{\rm sat}}(E) = \frac{1}{\sqrt{8}\pi^2}\frac{\dd}{\dd E}\int_0^E\frac{\dd\Phitot}{\sqrt{E-\Phitot}}\frac{\dd\rho_{i, {\rm sat}}}{\dd\Phitot},
\end{equation}
where $i$ labels the stellar ($i=\star$) or the dark matter ($i={\rm dm}$) DFs, while $\Phitot$ is the total potential (stars and dark matter). Note that $\PhistS$ is found numerically from the Poisson equation after computing $\rhostS$ from equation~(\ref{for:sat8760st}) via the Abel Inversion
\begin{equation}\label{for:abel}
    \rhostS(r) = -\frac{1}{\pi}\int_r^{+\infty}\frac{\dd\dlosS}{\dd t} \frac{\dd t}{\sqrt{t^2-r^2}}.
\end{equation}

As for NGC 5238, we may expect UGC 8760 to be orbited by one satellite only. We, thus, fixed the parameters of the satellite of UGC 8760 to $\MstS=6\times10^5\Msun$, $m=0.75$ and $\Reff=0.15\kpc$, while $\MdmS=1.5\times10^8\Msun$ and $\admS=0.5\kpc$. The chosen sizes and dark matter mass are such to mimic those of local dwarf elliptical or dwarf spheroidal galaxies (see, for instance Tucana, Carina, Andromeda XVI or Andromeda XVIII; \citealt{McConnachie2012}). The selected satellite halo parameters are, in any case, consistent with those one would obtain based on cosmological considerations, using the same approach employed in NGC 5238.

Differently from NGC 5238, in the case of UGC 8760, it is possible to make a rather rudimentary estimation of its satellite stellar mass based on the luminosity of the substructure. Considering the uncertainty regarding the actual extent of the overdensity observed to the north, and after estimating the background in that region, we obtain ${\rm M_r}=-9.05$ and ${\rm M_g}=-8.2$ (at the adopted distance) as r-band and g-band absolute magnitudes. After converting the g/r magnitueds into V-band magnitude (Lupton 2005)\footnote{\url{https://www.sdss3.org/dr9/algorithms/sdssUBVRITransform.php\#Lupton2005}.}, considering mass-to-light ratios compatible with those of an old stellar populations (2-6), and marginalizing over a $\sim0.5$ mag error, the stellar mass can well be within $(0.2-1.5)\times10^6\Msun$, in agreement with the value we adopted.

\subsection{Initial conditions}
\label{sec:ICs}

All the simulations are performed with the publicly available version of the hydrodynamical $N$-body code AREPO \citep{Springel2010,Weinberger2020}. AREPO blends the advantages of both Lagrangian and Eulerian hydrodynamics by solving Euler equations on an unstructured Voronoi mesh that moves with the fluid flow. The gas moving mesh is integrated with a particle-mesh algorithm and an oct-tree approach \citep{Barnes1986} for solving the Poisson equation and calculating accelerations for both collisional and collisionless particles. We adopt a mesh refinement algorithm to refine (de-refine) the gas cells when they become more (less) massive than a target reference mass. This choice allows to sample high density regions with a larger number of gas cells.

The dark matter and stars of NGC 5238 and the dark matter of its satellite have a mass particle of $\mpart=700\Msun$, while the satellite's stellar component is sampled with a particle mass of $\mpart=350\Msun$. Similarly, for all components of UGC 8760 (stars, dark matter, and gas), we adopt a particle mass of $\mpart=350\Msun$. For the satellite, we assume a dark matter particle mass of $\mpart=350\Msun$, while we adopt a lower mass of $117\Msun$ for its stellar component to ensure adequate sampling. The dark matter halos are sampled out to 10 times the virial radius of the associated NFW halo ($=53a$), where 98\% of the total mass is contained. The softening of each component is such that the maximum force between the particles of that component should not be larger than the mean-field strength they generate. Table~\ref{tab:params2} summarizes the number of particles for each component and the softening lengths used in the simulations. Since the galaxy models of NGC 5238, UGC 8760 and their satellites have been sampled via Maxwellian and epicyclic approximations, they are not guaranteed to be strictly in equilibrium. Therefore, we let them evolve in isolation for $6\Gyr$ to make them shift toward equilibrium. We also ensured that none of the galaxies developed bars due to disc instabilities when evolved in isolation. This approach is similar to that used in \cite{Pascale2021,Pascale2022}. The outcomes of these equilibrium simulations are then used as ICs for the simulations where the galaxies interact. 

In the simulations where we try to reproduce the interaction between hosts and satellites, the centers of mass of NGC 5238 and UGC 8760 are placed at the center of their respective simulation boxes. In the case of NGC 5238, the disc of the galaxy is inclined by $55^\circ$ with respect to the positive $z$-axis. Its satellite is at the initial position $\xyz=(15\kpc,0,0)$ with initial velocity $\vxyz=(-5\kms,0,0)$, i.e. on a radial orbit that will cross the host. These ICs are tailored to increase the strength of the gravitational interaction between host and satellite, with the satellite crossing NGC 5238 in the north-south direction, as the direction of the observed substructures. The satellite's disc lies on the $xy$-plane and rotates clockwise. As we discuss in Section~\ref{sec:res}, in this configuration ram pressure completely removes the gas from the satellite upon its first passage. Thus, specifying whether the satellite is a disc or a spheroid becomes less critical in NGC 5238 rather than in UGC 8760. It is only essential that, post-interaction, the satellite’s stellar component either falls below the surface brightness limit or undergoes sufficient disruption, which is easily accomplished given its stellar mass and size.

In the case of UGC 8760, its disc is inclined by $75^\circ$ counterclockwise with respect to the positive $z$-axis, with the $x$-axis lying in the plane of the disc. Its satellite is positioned at $\xyz=(5\kpc,0,0)$, with initial velocity $\vxyz=(0,25\kms,0)$, i.e. on an orbit that would be circular if we did not account for dynamical friction. The discs of UGC 8760 and NGC 5238 would rotate counterclockwise if they belonged to the $xy$-plane. Note that $55\deg$ and $75\deg$ are just how much NGC 5238 and UGC 8760 discS are inclined with respect to the $xy$-plane in the ICs. They would correspond to the formal discs inclinations only if the $y$-axis is assumed as line-of-sight.

In all the simulations we set the target reference mass to $700\Msun$. Note that due to mesh refinement, the initial total number of gas cells in the simulations reproducing the interactions is not the same as in the ICs of the simulations evolved in isolation.

\begin{figure*}[ht!]
    \centering
    \includegraphics[width=1.\hsize]{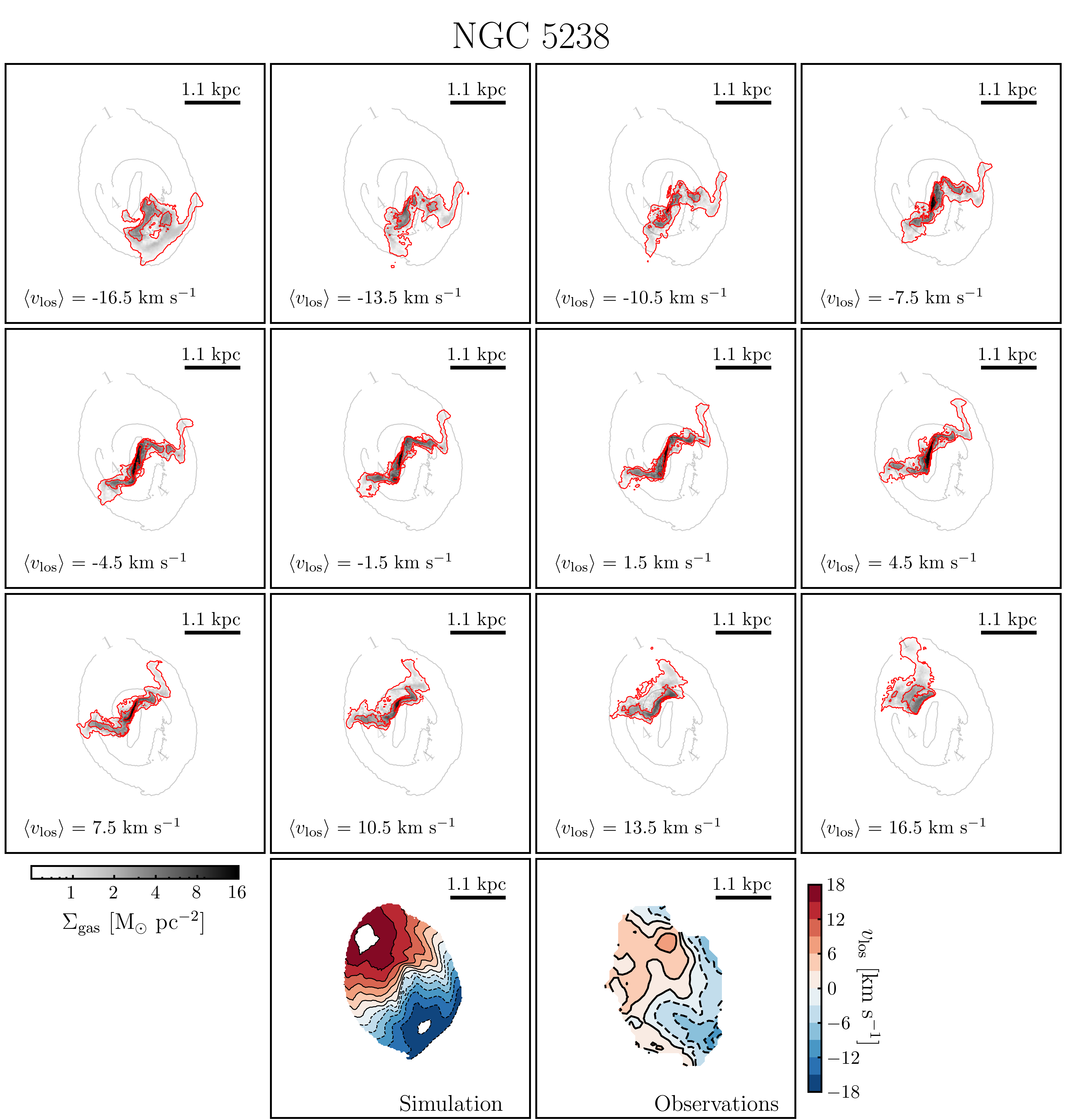}
    \caption{Channel maps of $\HI$ gas distribution of NGC 5238 computed from the reference simulations of Section~\ref{sec:res}. The channels width is $3\kms$, with the average channel velocity reported in each panel. The red curves show $\HI$ column densities of 0.5, 2 and 8 $\Msun\pc^{-2}$ while, in black, the same 0.5, 2 and 8 $\Msun\pc^{-2}$ isodensities but computed from the total gas distribution. The bottom panels show the overall gas-mass weighted line-of-sight velocity field in the simulation (left) and in the VLA observation \citep{Cannon2016}: blue colors and negative velocities correspond to the disc approaching arm, while red colors and positive velocities to the disc receding arm. The velocity field in the simulation extends down to column densities of 0.5 $\Msun\pc^{-2}$. The panel size is $5.5\kpc\times5.5\kpc$.}
    \label{fig:ugc5238vchannel}
\end{figure*}

\section{Results}
\label{sec:res}

Here, we present the results of these two simulations that accurately replicate the observable characteristics of the target galaxies. The specific cases of NGC 5238 and UGC 8760 are discussed separately in Sections~\ref{subsec:res5238} and ~\ref{subsec:res8760}, respectively. 

The sequence of panels in Fig.~\ref{fig:8760mock} shows a selection of surface brightness maps from different snapshots of the simulations of NGC 5238 (top) and UGC 8760 (bottom). The left-most column shows the ICs of the simulations, the second and third columns are subsequent snapshots, corresponding to $0.67\Gyr$ and $0.88\Gyr$ for NGC 5238, and $0.39\Gyr$ and $0.69\Gyr$ for UGC 8760. The surface brightness maps in the third column are meant to reproduce the LBT observations displayed, for a comparison, in the rightmost panels. In the mock surface brightness maps, the line-of-sight is obtained by rotating the systems counterclockwise first about the $x$-axis by an angle $\alpha$, then about the $z$-axis by an angle $\gamma$ to align the position angles to those of the observations. For NGC 5238 $(\alpha,\gamma)=(165^\circ,110^\circ)$, for UGC 8760 $(\alpha,\gamma)=(175^\circ,120^\circ)$. These rotations correspond, respectively, to disc inclinations of $i\simeq40^\circ$ and $i\simeq70^\circ$ w.r.t. the line-of-sight. In the simulated surface brightness maps, we converted mass into LBT g-band luminosity assuming constant mass-to-light ratios of $(M/L_g)/(M/L_{\rm bol})_\odot=3.2$ for NGC 5238 and its satellite, and $(M/L_g)/(M/L_{\rm bol})_\odot=3.2$ and 6 for UGC 8760 and its satellite. Here, $(M/L_g)/(M/L_{\rm bol})_\odot=3.2$ is the mass-to-light ratio corresponding to a $5\Gyr$ old stellar population with metallicity $Z=0.01$. This is a reasonable assumption for our simulated gas-rich dwarf galaxies because, according to numerous studies based on resolved-star CMD modeling \citep[e.g.; ][]{Tolstoy2009}, this type of systems is known to have experienced a continuous star formation since $10\Gyr$ ago (or earlier) to the present epoch, but we can exclude from our LBT photometry the presence of luminous stars younger than 1 Gyr in the extended low surface brightness galaxy outskirts that we intend to reproduce. On the other hand, $(M/L_g)/(M/L_{\rm bol})_\odot=6$, as expected for a 10 Gyr old stellar population with the same metallicity, is more appropriate for the satellite of UGC 8760 that we model as a gas-free spheroidal galaxy. Since the simulations do not model star formation, star particles cannot trace the dynamics of young stars. Therefore, by construction, the mock surface brightness maps shown in Fig.~\ref{fig:8760mock} cannot reproduce the brightest central parts of the galaxies where most of the contribution to light comes from young stars (see Fig.~\ref{fig:rgbstars}). In Fig.~\ref{fig:8760mock}, we used the same pixel size ($0.225\times0.225\asecq$) of the LBC camera and imposed a surface brightness limit of $\simeq30\magn\asec^{-2}$ as in the observations. The maps were smoothed with a Gaussian kernel of $0.45\asec$ ($\sim1\asec$ FWHM) dispersion to account for typical seeing during the observations. To help the visual comparison between data and simulations, the physical scales in the third and forth columns panels are the same.

\subsection{NGC 5238}
\label{subsec:res5238}

In the NGC 5238 simulation, the satellite galaxy starts from a significantly large distance, with a low initial velocity, aimed at reproducing a purely radial interaction (see top-left panel of Fig.~\ref{fig:8760mock}). Unlike the case of UGC 8760, these ICs are designed to increase the strength of the interaction between the two galaxies, a choice motivated by the pronounced irregularities observed to the north and south of the galaxy. After $0.67\Gyr$, the satellite galaxy has entirely crossed NGC 5238 once, and this passage is indeed highly disruptive for both the satellite system and the host. On NGC 5238 the gravitational effect is such to significantly distort its disc. The satellite, instead, loses coherence to the point that its surface brightness reaches the detection threshold, making it extremely faint. To aid in identifying the satellite galaxy, we display its stellar particles in blue in Fig.~\ref{fig:8760mock} (top row, third panel from left).

One significant consequence of this interaction is the formation of a predominant spiral arms pattern. The formation of spiral arms induced by interactions and mergers has been extensively explored through numerical simulations \cite[e.g. see][]{Gerin1990,Lang2014,Peschken2019} and they are recognized as a key mechanism contributing to the development of spiral arms within galaxies. Generally, tidally induced spiral arms are regarded as transient phenomena, with their longevity depending upon the mass ratio of the interaction, as discussed in \cite{Oh2008a} and \cite{Struck2011}. In cases where the mass ratio is 1:10, such as in the reference simulation of NGC 5238, these spiral arms can persist for several billion years. In our simulated scenario, it is intriguing to note that the substructures located to the north and south of NGC 5238 are not comprised of stars originating from the satellite itself. Instead, these substructures manifest as extensions of the spiral pattern induced by the interaction. Particularly noteworthy is the southern structure, which is more faithfully reproduced in the g-band image by the interaction. The northern 'umbrella'-like pattern, appears, instead, less accurately replicated being less extended in the simulation than in the observation. We do not consider this discrepancy in the replication of northern substructure a major concern. Indeed, as it can be noted from the RGB map in Fig.~\ref{fig:rgbstars}, the 'umbrella' feature is less dense than the southern structure and likely more noticeable in the g-band map due to the presence of background galaxies which are removed, instead, when computing the RGB star counts map. 

In the observations of NGC 5238 the spiral pattern is scarcely detectable within the central region of the galaxy. Nonetheless, it is crucial to emphasize, again, that the simulations do not incorporate star formation and its associated feedback. Consequently, the stellar particles primarily represent an old stellar population (prominently visible in the outer regions; Section~\ref{sec:ssh_data}) and they may not be very representative of the central regions, where the younger stellar component and the star forming gas predominantly influence the morphology of the galaxy \citep{Cannon2016}.

As anticipated, the satellite is scarcely detectable (see Fig.~\ref{fig:8760mock}, third column, top panel), since it undergoes an almost complete disintegration after less than an orbital revolution. This results in a substantial decrease in surface brightness, to the extent that it falls almost below the surface brightness threshold (average surface brightness $29.5\asec^2$). As we already mentioned in Section~\ref{sec:ICs}, the distinction between whether the satellite is a disk or a spheroid also becomes of minor importance, since the observed features are extensions of the induced spiral pattern and since the satellite has been depleted of gas as a result of ram pressure.

As a comparison, in the LBT surface brightness maps of Fig.~\ref{fig:8760mock} (rightmost panel) we superimposed levels of $\HI$ isodensities taken from the VLA observations from \cite{Cannon2016} for NGC 5238 and from \cite{Begum2008} for UGC 8760. The same quantities have been computed from the simulations and shown in the first three columns. The isodensity levels correspond to 1, 2, 4 and 8 $\Msun\pc^{-2}$. As noted by \cite{Cannon2016}, the outer $\HI$ disc displays a few asymmetries, with its morphological major axis elongated from northeast to southwest. The high-column density $\HI$ gas exhibits a crescent-shaped morphology, with the maximum of the $\HI$ surface density offset from the central optical peak by approximately $300\pc$ to the northeast, and not co-spatial with the brightest UV and $\Halpha$ emission. In the simulation, the gas distribution appears to be smoother than in the VLA observations by \cite{Cannon2016} and the offset is not reproduced. This discrepancy probably arises from the fact that our simulations are inherently adiabatic, thereby lacking the inclusion of star formation and associated feedback processes, preventing the faithful replication of distortions associated with these phenomena. Nevertheless, following the interaction between the satellite and the galaxy, we manage to reasonably replicate the spatial gas distribution. The overall extent of the gas distribution is, nevertheless, well reproduced although the outer $\HI$ gradient tends to be somewhat steeper in the observations than in the simulation post-interaction, as it can be appreciated looking at the different outer density of $\HI$ iso-contours in Fig.~\ref{fig:8760mock} (third and forth panels). The $\HI$ distribution in the simulation aligns well with recent, unpublished VLA observations, confirming a confined gas distribution that is less extended compared to the spatial distribution of stars. A detailed analysis of the $\HI$ distribution and its kinematics, based on these novel VLA observations, will be presented in an upcoming paper (Cannon et al., in preparation).

In Fig.~\ref{fig:ugc5238vchannel} we show the $\HI$ density in velocity channels computed from the simulations together with the overall $\HI$ velocity map compared with the analogous quantity computed from the simulation. The observed $\HI$ velocity field has significant asymmetries throughout the disc, featuring an S-shaped velocity profile that has been suggested by \cite{Cannon2016} as a potential indicator of a warp in the disc. In the averaged velocity field from the simulation (Fig.~\ref{fig:ugc5238vchannel}, bottom-left panel), this S-shaped feature is fairly well replicated, especially in the central regions of the galaxy. As observed by \cite{Cannon2016}, the galaxy channel maps reveal emission over approximately $40\kms$, peaking at $24\kms$ at $1.3\kpc$. The average velocity field in the simulation exhibits a larger amplitude, although the emission by channels of velocity closely reproduces the galaxy's velocity gradient and the galaxy's emission across its entire extent. The higher amplitude of the average velocity field in the simulation must not be regarded as a limitation of the simulation, as the velocity field amplitude on the plane of the sky is heavily influenced by the line-of-sight. A small variation around the assumed line of sight results in a consistent reduction in the velocity field amplitude, but keeping the emission by channel less affected. Adopting $i\simeq40\deg$, as in our case, simply ensures a good compromise in terms of representing both the structural and kinematic components. 

\subsection{UGC 8760}
\label{subsec:res8760}


In the case of UGC 8760, to facilitate the gradual approach of the satellite galaxy towards the host in a reasonable amount of time, we leverage dynamical friction which, according to our calculation, acts on a timescale that is, approximately, $\tfric\simeq0.5\Gyr$.\footnote{We computed the dynamical friction time as in equation (8.13) of \cite{BinneyTremaine2008}, with $\tcross=1.22\Gyr$ the crossing time estimated as the time needed to complete a circular orbit of radius $d=5\kpc$ with a velocity of $25\kms$, assuming a Coulomb logarithm $\ln\Lambda=15$.} In addition to that, the truncation radius of the satellite, that can be approximated as $\rt=(\frac{\Msat}{\Mhost})^{\frac{1}{3}}d\simeq2\kpc$ \citep{BinneyTremaine2008}, with $\Msat$, $\Mhost$, and $d$, respectively the total mass of the satellite, the mass of the host, and the distance between the two centers at the beginning of the simulation. The fact that the satellite's $\rt$ is larger than its size (Table~\ref{tab:params}) and way shorter than the initial distance $d=5\kpc$ ensures that any gravitational perturbation provided by UGC 8760 is progressively accounted for in the simulation and that it is not an artifact of the initial set up.  

\begin{figure}
    \centering
    \includegraphics[width=1\hsize]{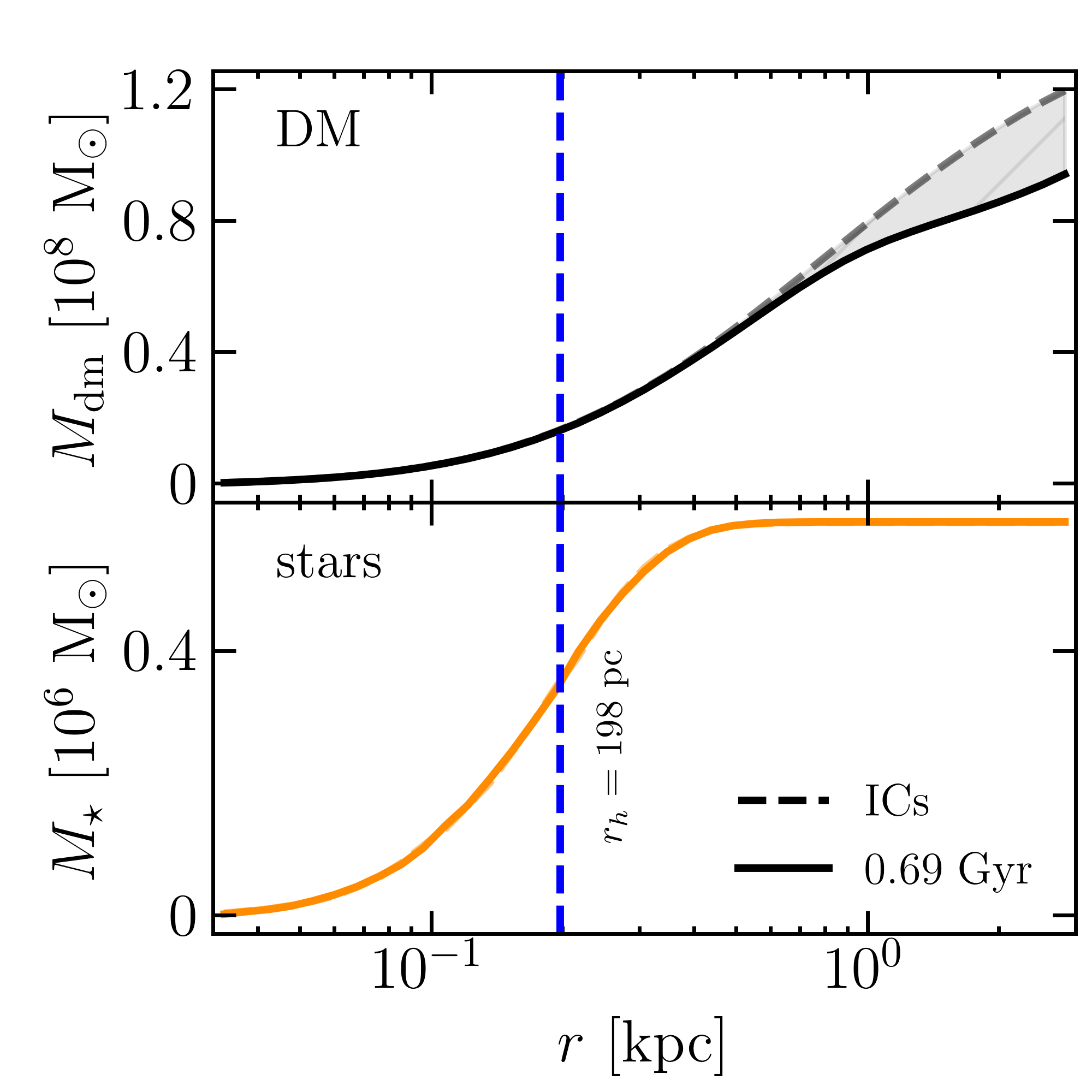}
    \caption{Top panel: dark matter mass distribution of the satellite of UGC 8760 as a function of the distance from the satellite center. The black-dashed line shows the mass distribution as in the ICs while the black solid after $t=0.69\Gyr$. Bottom panel: same as in the top panel but showing the mass distribution of the stellar component on the satellite. The dashed blue line marks the position of the stellar component half-mass radius. The stellar mass distribution does not change from the ICs to $0.69\Gyr$.}
    \label{fig:8760satdens}
\end{figure}

\begin{figure*}
    \centering
    \includegraphics[width=1.\hsize]{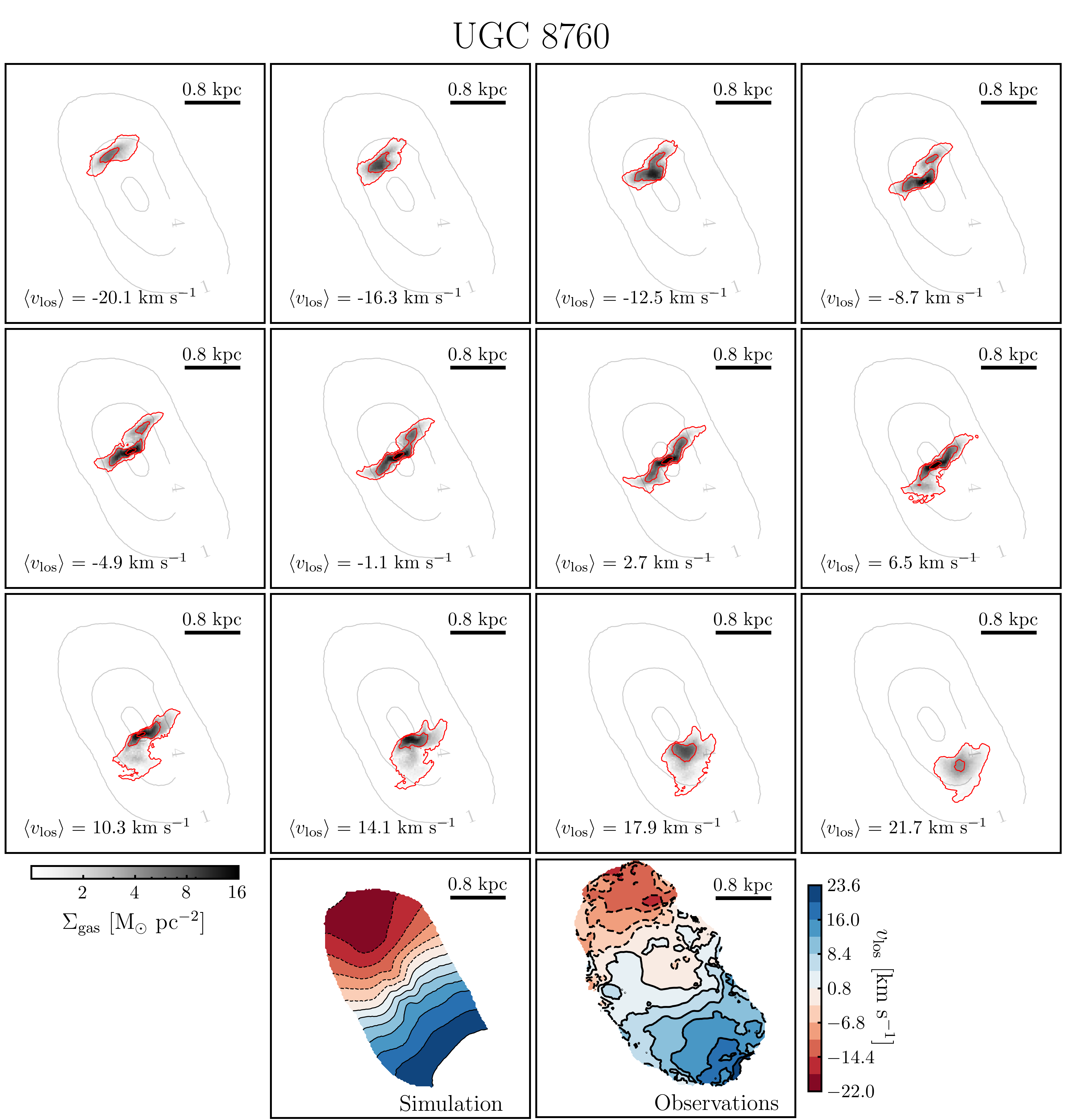}
    \caption{Same as in Fig.~\ref{fig:ugc5238vchannel} but showing the $\HI$ channel map of UGC 8760. Here, the channel width is $3.8\kms$, the box size is $4\kpc\times4\kpc$, while the observations are from \citet{Begum2008}.}
    \label{fig:ugc8760vchannel}
\end{figure*}

As the simulation starts, the satellite galaxy sinks towards the center in a time compatible with $\tfric$: after half an orbit, taking less than $0.39\Gyr$, the average distance between the host and the satellite has decreased to $2.4\kpc$ (first row, second panel of Fig.~\ref{fig:8760mock}), while it is $1.6\kpc$ after $0.69\Gyr$ (first row, third panel of Fig.~\ref{fig:8760mock}). At $t=0.69\Gyr$ the satellite galaxy is, in projection, approximately at the same distance as the old stellar overdensity observed in UGC 8760. In the simulation, the satellite is stripped of a considerable amount of dark matter in the outer regions. For a comparison, Fig.~\ref{fig:8760satdens} shows the enclosed mass distributions of dark matter (top) and stars (bottom) of the satellite at the beginning of the simulation and after $0.69\Gyr$. Here, the mass outside $3\kpc$ is significantly reduced from its initial value, with a fractional mass loss of approximately 21\%. However, despite the decreased dynamical mass, the stellar component, which is very compact and concentrated at the center of the satellite, maintains a smooth and regular density distribution, remaining nearly unchanged compared to the initial one, despite its proximity to the host. In this configuration, the orbit of the satellite galaxy is almost co-planar with the plane of the sky. Therefore, the residual average line-of-sight velocity of the satellite with respect to the main galaxy is low, about $6\kms$.

The galaxy $\HI$ density distribution is shown in Fig.~\ref{fig:8760mock} by the red contours. In the VLA observations of UGC 8760, the outer $\HI$ gas disc distribution is slightly bent, seemingly symmetric towards the northwest, with the outer distribution displaying a box-like structure. In the simulation, the UGC 8760 gas disc undergoes mild distortions in the outer regions due to the close passage of its satellite. However, after $\simeq0.69\Gyr$, although the UGC 8760 gas distribution develops a similar outer boxy-like pattern, the bent structure is more prominently emphasized by the stellar distribution rather than the $\HI$ in the simulation. In the simulation, these stars serve to bridge the gap between UGC 8760 and the satellite. This bridge is evident in the RGB count map (Fig.~\ref{fig:rgbstars}, right panel) and in the simulations, but is not detectable in the g-band image. The reason for the absence of this star bridge in the g-band image is primarily attributed to its low surface brightness, coupled with significant contamination from background galaxies, which is absent in the mock image. RGB maps, on the other hand, can achieve deeper surface brightness levels as they enable the isolation of stars and the removal of a large portion of background galaxies.

Fig.~\ref{fig:ugc8760vchannel} shows the same channel maps of $\HI$ densities shown in
Fig.~\ref{fig:ugc5238vchannel}, but for UGC 8760. The simulation effectively replicates the velocity field derived from $\HI$ observations of the galaxy, particularly in terms of the overall gradient and the position of the approaching and receding arms. However, as for NGC 5238, it is important to note that the observed velocity field exhibits a certain degree of perturbation and granularity that cannot be recreated by adiabatic simulations that do not account for stellar feedback phenomena. In this respect, the mock velocity field computed from the simulation does appear much smoother and regular than the observed one.


\subsection{Additional simulations}
\label{subsec:newsim}

\begin{figure*}
    \centering
    \includegraphics[width=1\hsize]{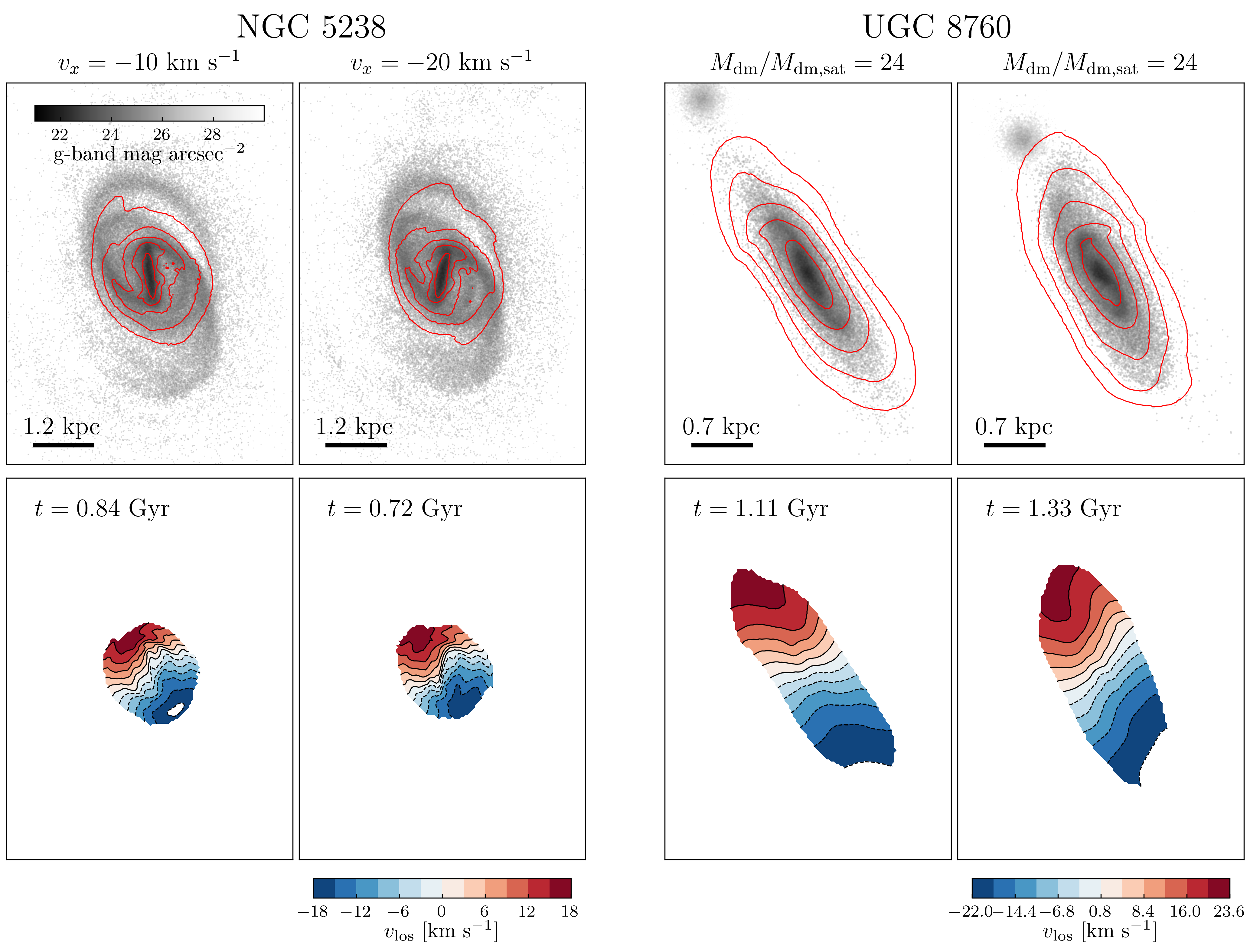}
    \caption{First two columns of panels: stellar surface brightness maps (top) and line-of-sight velocity field (bottom) from two additional simulations of NGC 5238 discussed in Section~\ref{subsec:newsim}. The columns, from left to right, show the simulations with initial velocities $\vx=-10\kms$ and $\vx=-20\kms$, respectively. The mock surface brightness and velocity field maps are obtained as in Fig.s~\ref{fig:8760mock} and \ref{fig:ugc5238vchannel}. Second two columns of panels: same as the first two columns, but for UGC 8760. In this case, instead of showing two different simulations, the columns show the same simulation but at different snapshots that both replicate the galaxy observed structure and kinematics. The merger mass ratio of the simulation is $\Mdm/\MdmS=24$.}
    \label{fig:newsim}
\end{figure*}

In this section, we briefly present the outcomes resulting from the exploration of additional parameters in the ICs of the simulations of NGC 5238 and UGC 8760. These parameters were varied to investigate how they might influence the simulations' ability to reproduce observational data. This sampling, however, covers only a small portion of the extensive parameter space, as a comprehensive exploration extends beyond the scope of this article.

In the case of NGC 5238, we have conducted simulations exploring the effects of varying the initial velocity of the satellite. Specifically, we tested velocities of $\vx=-10\kms$ and $\vx=-20\kms$, while keeping all the other orbital and galaxy models parameters the same. The two left-hand columns of panels of Fig.~\ref{fig:newsim} show mock surface brightness maps of the NGC 5238 system from the two aforementioned simulations, corresponding to snapshots that reproduce the observations. The line of sight is approximately the same as that adopted in the mock images from the reference simulation, while the time from the beginning of the simulation is $t=0.84\Gyr$ and $t=0.72\Gyr$, from left to right. Due to the varying initial velocities, the corresponding time also differs, being slightly greater for lower initial velocities. The maps also show $\HI$ isodensity contours in red, while the lower panels the corresponding line-of-sight velocity field. As can be readily observed, the outcome of the simulations closely resembles that of the reference simulation. Despite the decrease in initial velocity, the satellite is nonetheless completely stripped of gas due to ram pressure. The perturbations induced in both the stellar and gaseous discs of NGC 5238 remain nearly unchanged, indicating little dependence on this parameter. Therefore, as long as the accretion occurs approximately on a radial orbit, the proposed type of interaction seems to reproduce quite faithfully the observed properties of the system.

As previously emphasized, these simulations do not represent the sole merger possibilities capable of reproducing the observed substructures in NGC 5238. As outlined in Section~\ref{sec:ICs}, the choice behind a radial interaction was motivated by the observed north-south directionality of the substructures, but it does not preclude other scenarios. For instance, it is reasonable to speculate that a flyby involving a satellite on a circular orbit, as in the UGC 8760 case, may similarly yield the disc of NGC 5238 to develop an outer spiral structure pattern ascribable to the north and south overdensities. From the parameters of the galaxy models presented in Table~\ref{tab:params}, it can be noted that, in many aspects, the UGC 8760 system is approximately a scaled down version of the NGC 5238 system, with approximately half of the dark matter mass in host and satellite, with a stellar disc extended by half, and four time less massive. Indeed, despite the unfavorable line of sight, the spiral pattern also emerges in the UGC 8760 system, as illustrated in Fig.~\ref{fig:8760mock}. Therefore, it is reasonable to speculate that a similar behavior may occur in the NGC 5238 system, provided appropriate scaling of the orbital ICs.

In the case of UGC 8760, the satellite's stellar-to-dynamical mass ratio is only 0.5\% within $3\kpc$, effectively rendering stars as mere tracers of the total potential. Consequently, reducing the satellite's stellar mass is unlikely to yield significant alterations in the simulation's outcome after $0.69\Gyr$, except for a decrease in luminosity, which can be compensated for by adjusting the mass-to-light ratio. As an alternative, we conducted simulations in which we maintained the same orbital parameters but varied the satellite's dark matter mass, and vice versa. When keeping the orbital parameters fixed (i.e. the satellite's initial position and velocity), the alteration in dark matter content was intended to explore interactions with different merger mass ratios. Specifically, in the first case we halved the dark matter mass, moving to a ratio $\Mdm/\MdmS=24$ instead of $\Mdm/\MdmS=12$ as in the reference simulation. Here, the gravitational perturbation of the satellite on the primary galaxy (UGC 8760) is significantly reduced. In the third and forth columns of panels of Fig.~\ref{fig:newsim} we show two snapshots of this simulation where the observations are fairly well reproduced.  Despite the substantially increased time required to reach UGC 8760, caused by the diminished mass and, thus, a less effective dynamic friction, the satellite's stellar component remains relatively unperturbed. Moreover, the perturbations experienced by the stellar and gas discs of UGC 8760 are much less significant as well. Nevertheless, also this simulation appears suitable to reproduce the observed structure and kinematics of UGC 8760. In the second case, we tested ICs with purely radial orbits (as in the case of NGC 5238, where the satellite crosses the galaxy's center, see Sections~\ref{subsubsec:5238} and ~\ref{subsec:res5238}), adopting two dark matter halos with varying amount of mass. Here, the gravitational interaction proves to be much more destructive to UGC 8760 and the coherence of its stellar and gas discs, potentially making such interactions far less likely. This exploration suggests that interactions featuring more circular orbits tend to foster gentler and less destructive outcome\footnote{A similar behavior is reported by \cite{Nipoti2012}, who find that, in minor mergers, the structural and dynamical properties of the main galaxy are less modified when the orbit has larger pericentric radius (see their figure 5)}, easily reproducing the observed structure of the UGC 8760 system, almost independently of the satellite dark matter mass within the range explored.

\section{Conclusions}
\label{sec:conc}

In this study, we presented the cases of NGC 5238 and UGC 8760, two dwarf galaxies observed as part of the SSH survey, exhibiting promising signs of recent interactions in deep LBT imaging. Stellar population studies based on these LBT observations revealed that the low-surface brightness and anomalous stellar substructures observed in the outskirts of the galaxies are predominantly composed of old stars, indicating limited association with sporadic and recent episodes of star formation. Through hydrodynamical $N$-body simulations presented in this work, we have shown that recent interactions
with less massive satellite galaxies can reproduce reasonably well the structural and kinematic characteristics
of both galaxies.

In both cases, the dwarf galaxies are represented by means of realistic ICs comprising of thick star and gas discs, embedded in a dominating dark matter halo, in agreement with state-of-the art measurements based on $\HI$ kinematics and stellar photometry. The simulations do not include star formation and feedback from supernovae explosions, implying that stellar particles trace only the kinematics of the old stellar populations of NGC 5238 and UGC 8760.

For the NGC 5238 system, the simulations suggest that a close interaction with a satellite galaxy  whose stellar mass is 50 times less than that of the host, and with a merger mass ratio 1:10, can reproduce some of the observed structural irregularities in the galaxy outskirts. In the scenario proposed, the overdensities of old stars observed to the north and south of the galaxy are mainly composed by stars belonging to the disc of NGC 5238, distorted by the close interaction with its satellite. Similarly, our simulations reveal that interactions with a small spheroid with a merger mass ratio of up to 1:12, provide a compelling explanation also for the presence of the low-surface brightness companion situated to the north of UGC 8760. In terms of gas kinematics, our simulations fairly replicate the distribution of $\HI$, as well as the overall velocity field in both cases. While a systematic exploration of the wide parameter space was not undertaken, the presented simulations suggest that the observed substructures emerge relatively rapidly after the simulation begins, potentially indicating a brief timescale for the interaction. Also, in the explored parameter space, in UGC 8760 we found that interactions with tangential orbit interactions produce more favorable outcomes compared to those with radial encounters, while radial interactions best reproduce the substructures of NGC 5238. 

Within the SSH survey, NGC 5238 and UGC 8760 complement the well-known cases of DDO 68 \citep{Annibali2016,Pascale2022} and NGC 3741 \citep{Annibali2022b}, and will be accompanied by additional dwarf galaxies that will be introduced and analyzed in details in an upcoming paper \cite{Sacchi2024}. All of these galaxies reside in low-density environments: within voids or on the periphery of well-known groups. Unlike dwarf galaxies in more densely populated regions, these isolated or low-density environments offer protection against disruptive interactions that, in a hierarchical scenario of structure formation, could otherwise occur with more massive hosts \citep{Deason2015}. \cite{Dooley2017} showed that for galaxies with a fixed stellar mass located relatively far from massive hosts like the Milky Way, the number of satellite galaxies orbiting around can be predicted using cosmological simulations. For instance, for a host with a stellar mass of $10^7-10^8\Msun$ like NGC 5238 and UGC 8760, there is a high probability of being orbited by at least one satellite system with a stellar mass of $\simeq10^5\Msun$, with the number of expected satellites increasing as we consider even smaller masses (up to three for a stellar mass of $10^4\Msun$). Indeed an increasing number of low-mass systems have been discovered in the past years \citep{Bellazzini2013,Erkal2020,Carlin2021,MartinezDelgado2021}. In conclusions, environments like those surrounding NGC 5238 and UGC 8760, which lack nearby massive galaxies, offer odds of survival for small satellites significantly improved. This is precisely the type of environment that would favor the presence of satellites of satellites. 

\begin{acknowledgements}
This paper is supported by the  Fondazione  ICSC , Spoke 3 Astrophysics and Cosmos Observations. National Recovery and Resilience Plan (Piano Nazionale di Ripresa e Resilienza, PNRR) Project ID CN\_00000013 "Italian Research Center for High-Performance Computing, Big Data and Quantum Computing"  funded by MUR Missione 4 Componente 2 Investimento 1.4: Potenziamento strutture di ricerca e creazione di "campioni nazionali di R\&S (M4C2-19 )" - Next Generation EU (NGEU).
\end{acknowledgements}

\bibliographystyle{aa}
\bibliography{paper}

\end{document}